\documentclass[twocolumn]{aastex61}

\newcommand\aastex{AAS\TeX}

\received{Aug 01, 2023}
\revised{Oct 20, 2023}
\accepted{Oct 22, 2023}

\shorttitle{\aastex\ manuscript article}
\shortauthors{Jiang et al.}

\begin{document}

\title{Observational evidence of a centi-parsec supermassive black-hole binary existing in the nearby galaxy M81}

\correspondingauthor{Wu Jiang \& Zhiqiang Shen}
\email{jiangwu@shao.ac.cn, zshen@shao.ac.cn}

\author[0000-0001-7369-3539]{Wu Jiang}
\affiliation{Shanghai Astronomical Observatory, 
 Chinese Academy of Sciences, 
 Shanghai 200030, China}
\affiliation{Key Laboratory of Radio Astronomy and Technology, Chinese Academy of Sciences, A20 Datun Road, Chaoyang District, Beijing, 100101, P. R. China}

\author[0000-0003-3540-8746]{Zhiqiang Shen}
\affiliation{Shanghai Astronomical Observatory, 
 Chinese Academy of Sciences, 
 Shanghai 200030, China}
\affiliation{Key Laboratory of Radio Astronomy and Technology, Chinese Academy of Sciences, A20 Datun Road, Chaoyang District, Beijing, 100101, P. R. China}
 
\author[0000-0003-3708-9611]{Ivan Mart\'i-Vidal}
\affiliation{Dpt. Astronomia i Astrof\'isica, Univ. Val\`encia,
C/Dr. Moliner 50, Agust\'in Escardino, E-46100 Burjassot, Spain}
\affiliation{Observatori Astron\`omic, Univ. Val\`encia,
C/Cat. Agust\'in Escardino 2,
E-16980 Paterna, Spain}
 
\author[0000-0002-5385-9586]{Zhen Yan}
\affiliation{Shanghai Astronomical Observatory, 
 Chinese Academy of Sciences, 
 Shanghai 200030, China}
 
\author[0000-0002-1923-227X]{Lei Huang}
\affiliation{Shanghai Astronomical Observatory, 
 Chinese Academy of Sciences, 
 Shanghai 200030, China}
\author[0000-0003-2492-1966]{Roman Gold}
\affiliation{CP3-Origins, University of Southern Denmark, Campusvej 55, DK-5230 Odense M, Denmark}
\author[0000-0002-7329-9344]{Ya-Ping Li}
\affiliation{Shanghai Astronomical Observatory, 
 Chinese Academy of Sciences, 
 Shanghai 200030, China}
 \author[0000-0001-9969-2091]{Fuguo Xie}
\affiliation{Shanghai Astronomical Observatory, 
 Chinese Academy of Sciences, 
 Shanghai 200030, China}
\author[0000-0002-7776-3159]{Noriyuki Kawaguchi}
\affiliation{National Astronomical Observatory of Japan, 
 2-21-1 Osawa, Mitaka, Tokyo 181-8588, Japan}



\begin{abstract}
Studying a centi-parsec supermassive black-hole binary would allow us to explore a new parameter space in active galactic nuclei and, besides, these objects are a potential source of gravitational waves. We report evidence that a supermassive black-hole binary with an orbital period of $\sim30$\,years may be resident in the nearby galactic nucleus M81. This orbital period and the known mass of M81 imply an orbital separation of $\sim0.02$ parsecs. The jet emanating from the primary black hole showed a short period of jet wobbling $\sim16.7$ years, superposing a long-term precession at a time-scale of several hundred years. Periodic radio and X-ray outbursts were also found twice per orbital period, which could be explained by a double-peaked mass-accretion-rate variation per binary orbit. If confirmed, M81 would be one of the closest supermassive black-hole binary candidate, providing a rare opportunity to study the ``final parsec problem".

\end{abstract}




\section{Introduction} \label{sec:Intro}
The nearby $\sim$3.96\,Mpc \citep{2007ApJ...668...924B} spiral galaxy M81 (NGC\,3031), harbors one of the nearest low luminosity active galactic nuclei \citep{2002A&A...392..53N, 2008ARA&A...46...475H}. The almost face-on aspect with an inclination angle to the line of sight (LOS) $14\pm2$ degrees of the host galaxy allows a clear and unobscured view of the nucleus. With a black hole mass $M\thicksim7\times10^7M_\odot$ determined spectroscopically \citep{2003AJ...125...1226D} with the Hubble Space Telescope (HST), an angular resolution of 1\,micro-arcsecond ({\textmu}as) corresponds to a linear size of 6 gravitational radius ($r_g=\frac{GM}{c^2}$). As seen with very long baseline interferometry (VLBI), the AGN in M81 (i.e., M81*) was identified with a compact core and a one-sided jet to the northeast \citep{1996ApJ...457...604B,2000ApJ...532...895B, 2008ApJ...681..905M}. The core-jet structure was confirmed by locating the core at multiple frequencies \citep{2004ApJ...615...173B}, and the frequency-size relation of the core was measured up to 88\,GHz \citep{2012A&A...537A..93R,2018ApJL...853..L14J}. The first series of VLBI observations on M81* were carried out in 1980s \citep{1982ApJ...262...556B}. Since the early 1990's, M81* has been frequently observed, mainly to monitor the evolution of the adjacent supernovae SN1993J \citep{1995Natur.373...44M,1997ApJ...486L..31M,2009A&A...505..927M,2000Sci...287..112B,2002ApJ...581..404B,2007ApJ...668...924B,2001ApJ...557..770B,2003ApJ...597..374B} and SN2008iz  \citep{2016A&A...593..A18K}, as well as to track the relative motions between M81 and M82. A long-term VLBI monitoring of M81* showed strong evidence of jet precession with a period of $\sim$7.27 years, it was also found a flare in the peak flux densities, which lasted from mid 1997 to mid 2001 \citep{2011A&A...533..A111M}. But the long-duration flare did not repeat in a late observation campaign in 2005 \citep{2008ApJ...681..905M}. The jet precession and its position angle (PA) drift were confirmed with new data in 2017-2019 \citep{2023A&A...672L...5V}, while it could not constrain the model well due to the sparse sampling in time domain.

In this paper, we collected all the archived VLA/VLBI and also our newly VLBI observational data from 1980 to 2022 at multifrequency bands, as well as the long-term X-ray monitoring data covering the same period. We found M81* underwent a periodic change of its jet PAs, with three main outbursts in both the radio and the X-ray. We propose a scenario involving a supermassive black hole binary in an orbit misaligned with the accretion disk that exists in M81*. The binary model fits the jet PAs well, and can explain the repeated outbursts. In Section 2 we summarize the observations and data reduction of M81*, we describe our results in Section 3 and the fitting by the proposed binary model in Section 4. Our discussions are presented in Section 5, followed conclusions in Section 6.

\section{observations and data reduction} \label{sec:obs}
\subsection{Radio}\label{subsec:VLBI}
\textbf{VLBI Observations} M81* has a flat or inverted spectrum at a flux density level of $\sim100$\,mJy. A high declination of $\sim70^\circ$ makes it 
circumpolar for most VLBI observatories in the Northern hemisphere, hence 
visible in common with all telescopes. 
We observed M81* with the Korean VLBI Network and VLBI Exploration of Radio Astrometry array (KaVA) in 2015 at 22 and 43\,GHz, the Very Long Baseline Array (VLBA) in 2016, 2018, 2019, 2021 and 2022 (the latest epoch had VLBA plus Tianma-65m and Sheshan-25m telescope in Shanghai) at multiple frequencies covering 5.0, 8.4, 22 and 43\,GHz. We also analyzed the archived VLBI observational data since 2005, which mainly included the VLBA and Effelsberg-100\,m (EF) telescope, some epochs had stations of European VLBI network, Greenbank (GB) Telescope and Very Large Array (VLA) joining. The 
VLBA, with additional inter-continental baselines, resolved M81* up to the ground-based highest resolution at each frequency. The usual 8 or 12 hour tracks for each epoch made the spatial-frequency sampling, known as UV coverage, rather complete for the image reconstruction. The published VLBI results of M81* in 1981 and from 1993 to 2005, with similar array configurations, were assembled for a complete understanding the periodic phenomenon. A summary of observations since 2005 is presented in Table \ref{tab:tab1}.

The VLBI data reduction followed standard procedures to the amplitude and phase calibrations in AIPS and Difmap, described in references \citep{2011A&A...533..A111M,2018ApJL...853..L14J}. Several iterations of phase self-calibration were executed in Difmap to make the final image, prior to an amplitude self-calibration with a solution interval of 10 minutes. The VLBI flux density measurements 
were estimated from the source core. Since the VLBA is an homogeneous array, the uncertainty of the amplitude calibration is $\sim$5\% \citep{2005AJ....130.2473K}. Hence, the VLBA data can be used as a template to constrain the initial source model in joint (i.e., global) observations.
In the case of KaVA, we observed the amplitude calibrator 4C\,39.25, whose flux density was $\sim$4.84\,Jy at 22\,GHz and $\sim$1.97\,Jy at 43\,GHz in our observations, almost at the same level as that in previous observations \citep{2014PASJ...66..103N}. While M81* showed a flux density of at least two times higher than its normal stage, the flux at 43\,GHz was even higher than that at 22\,GHz, indicating radio outbursts during these periods. 

\textbf{VLA Observations} The archived observational data of the interferometer VLA at 5.0, 8.4 and 15 GHz from 1980 to 1991 were re-analyzed with AIPS and identified an outburst around the middle of 1985. The VLA data around 1999 and 2015 were added for increasing the cadences during outburst phases. The 2014-2015 data was utilizing the CASA software package \citep{2022PASP..134k4501C}, version 5.5.0. The origin of CASA lies in the AIPS++ project and it is better suited for processing the datasets of 2014-2015 observations. The visibilities were examined for radio frequency interference, which primarily affected the low frequency bands. Standard reduction techniques were used to transfer the flux density scale from the flux density calibrator to phase calibrator and M81*, and then bandpass, gain calibrations, and primary beam correction (if necessary) were done. A Gaussian centered at the phase center was then fit to the visibilities. A 10\% error was added to the statistical fit error, to account for the uncertainty in the absolute flux density. The results are presented in Figure \ref{fig:fig1}B and Table \ref{tab:tab1}. 

\subsection{X-ray}\label{subsec:X-ray}
We analyzed the soft X-ray emission from the nuclear source of M81* from 2005 to 2022, using the archival observations of the X-ray Telescope (XRT) on board the Neil Gehrels {\em Swift} Observatory. The cleaned events were reduced by the tool {\tt xrtpipeline}. The source spectra were extracted from a circle region with a radius of 30 pixels, for photon counting (PC) mode, and a box region with a width of 60 pixels and a height of 20 pixels, for window timing (WT) mode. For the observations suffering pile-up effects (count rate $>$0.6 c/s for PC mode), we excluded the central pixels determined by using the updated XRT point spread function (PSF) \footnote{\url https://www.swift.ac.uk/analysis/xrt/pileup.php}. The background spectra were extracted from an annulus region with inner radius of 180 pixels and outer radius of 200 pixels, to avoid other point sources in the {\em Swift}/XRT filed of view.

The X-ray spectra of {\em Swift}/XRT were fitted with a model {\tt tbabs*(powerlaw)} in the energy range of 2--10 keV, where the column density $n_\mathrm{H}$ is 
fixed as the Galactic value $7.22\times10^{20}$ cm$^{-2}$ towards to the direction\footnote{\url{https://www.swift.ac.uk/analysis/nhtot/index.php}}. The X-ray spectral fitting was performed with Bayesian X-ray Analysis \citep[BXA,][]{2014A&A...564A.125B} within {\tt PyXspec} \citep{2021ascl.soft01014G}. Then, we used the convolution model {\tt cflux} to calculate the flux in the energy range 2--10 keV. We found a negative correlation between the photon index and X-ray flux, which was fitted with a function as $\Gamma = -0.55\times \log(F_\mathrm{2-10 keV}) - 4.25$, implemented in the software package \texttt{UltraNest} \citep{2016S&C....26..383B,2019PASP..131j8005B,2021JOSS....6.3001B}. 

In order to investigate the long term variability of M81*, we also used the flux measurements of other X-ray instruments in the literature. \citet{1982ApJ...257L..51E} reported the 2--10 keV X-ray flux of two {\em Einstein} observations during 1978--1979 (\autoref{fig:fig1}). \citet{1993ApJ...418..644P} reported the X-ray flux in 2--10 keV of $Einstein$, $EXOSAT$, $Ginga$, $ROSAT$ and $BBXRT$ (\autoref{fig:fig1}). \citet{2004ApJ...601..831L} reported the X-ray flux in 0.5 --2.4\,keV for more than twenty years, including the X-ray mission {\em ROSAT},{\em ASCA}, {\em Chandra} and {\em XMM-Newton}. In order to convert the X-ray flux into 2--10 keV, we used the photon index 1.79 for {\em ROSAT} and {\em ASCA}, and the reported photon index of {\em Chandra} and {\em XMM-Newton} in \citet{2004ApJ...601..831L}. 

\section{Results} \label{sec:results} 
\subsection{Position angle}\label{subsec:PA} 
\begin{figure}[h!]
\centering
\includegraphics[width=3.4in]{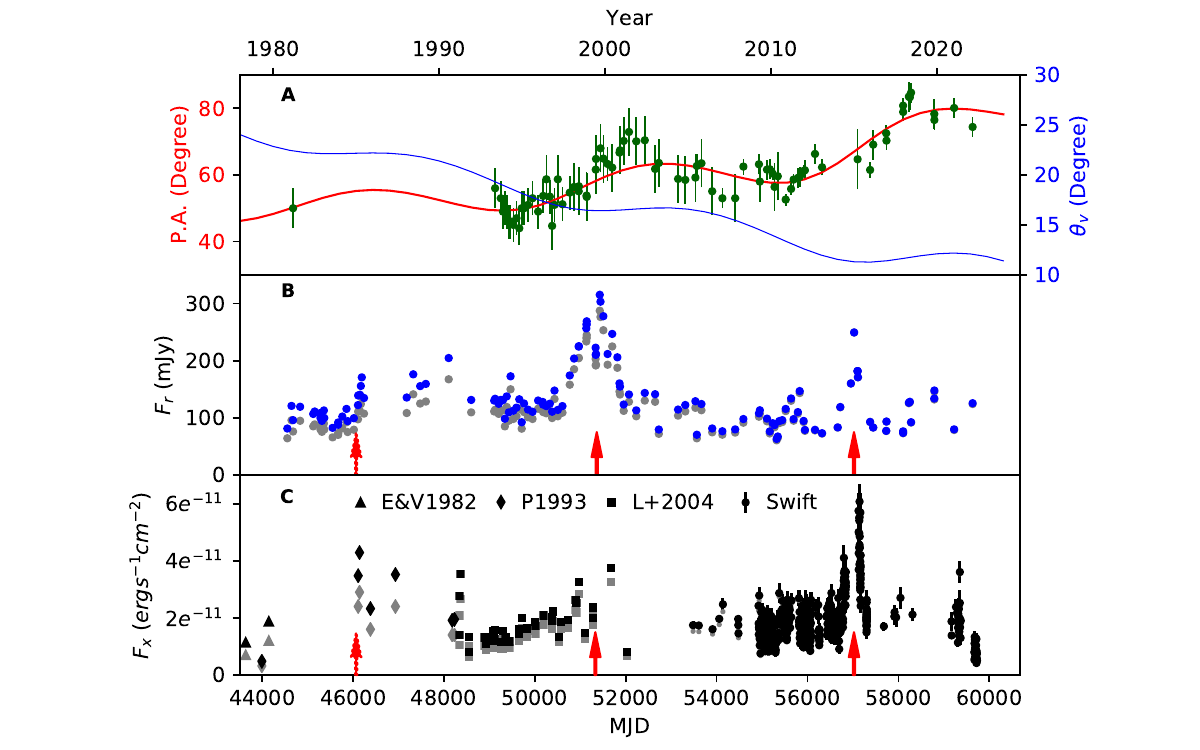}
\includegraphics[width=3.4in]{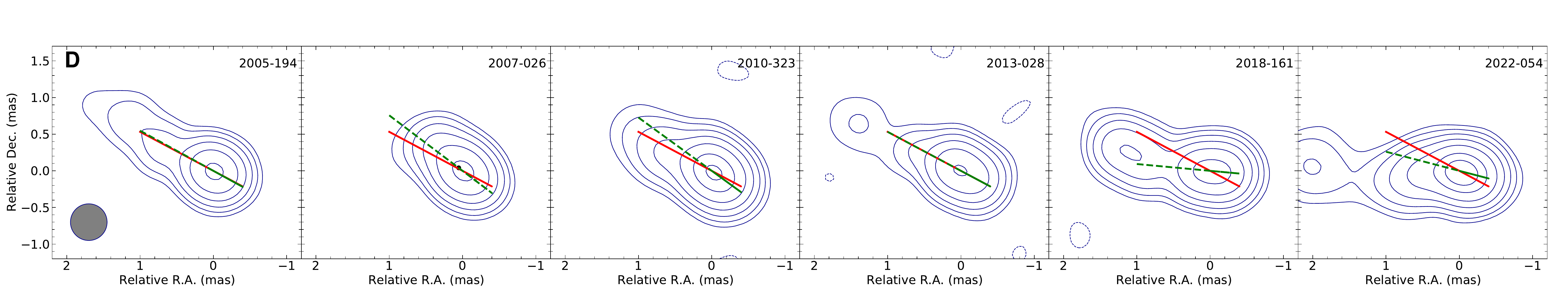}
\caption{Radio and X-ray variations of M81*  from 1980 to 2022. a) Jet position angles observed at multifrequencies and assembled at 8.4\,GHz. The solid curves are the Bayesian fitting to the SMBHB model and results of evolution of PA (red) and viewing angle (blue). b) Radio light curve formed assembled at 8.4\,GHz (grey: original; blue: Doppler beaming corrected). c) X-ray (2-10\,keV) light curve (grey: original; dark: Doppler beaming corrected). d) 6 epoch VLBI images at 8.4\,GHz from 2005 to 2022, all the uniform-weighted interferometry beams shown in bottom left corner are restored to a normal beam size (0.5\,mas$\times$0.5\,mas), which is the inter-continental interferometry including VLBA and EF or SH at 8.4\,GHz. The contour levels are 0.3$\times$(-1, 5, 10, 20, 40, 80,160, 240) mJy\,beam$^{-1}$. The lowest level is about three times root-mean-square of noise in the image. The red line indicates the projection on the sky of the rotation axis of M81 galaxy at $62^\circ$ \citep{1982ApJ...262...556B}; the dashed green lines indicate the position angles of jet in the center region from the results of model-fitting. \label{fig:fig1}}
\end{figure}

The compact core region of M81* at all frequencies is elongated in the direction of the jet extension and shows a flat or inverted spectrum \citep{1996ApJ...457...604B,2000ApJ...532...895B,2011A&A...533..A111M, 2018ApJL...853..L14J}. 
Hence, we fitted the VLBI calibrated visibility data with elliptical Gaussian models. The radio emission from the core could be well fitted by a single Gaussian \citep{2000ApJ...532...895B,2011A&A...533..A111M,2018ApJL...853..L14J}. The jet orientation in this model corresponds to the orientation of the Gaussian main axis. A similar approach has been applied by the team of Monitoring of Jets in Active Galactic Nuclei with VLBA Experiments and demonstrated to be a robust way to derive the core parameters \citep{2005AJ....130.1389L,2005AJ....130.2473K}. Besides this model of an elliptical Gaussian (plus an extended circular Gaussian, to account for the more distant jet emission), we also fitted the data using two close-by circular Gaussians, which resulted in fits with a similar quality. For this alternative model, the PA of the jet can be estimated from the orientation between the two Gaussian components.
The PAs derived from the elliptical-Gaussian fitting are consistent with those 
from the two circular-Gaussian components. 
We adopted the PAs estimated from the elliptical Gaussians from 1993 to 2005 \citep{2000ApJ...532...895B,2011A&A...533..A111M}, to cover another radio and X-ray outbursts. While the practical PA errors are three times of the norminal errors in their papers. In the simultaneous multi-frequency observations, the PAs at 5.0, 15, 22, and 43\,GHz were different to those observed at 8.4\,GHz by $1.9\pm3.3$, $-1.1\pm1.6$, $-1.3\pm0.7$, and $-3.1\pm1.4$ degrees, respectively, implying a slight frequency-dependent jet orientation, likely related to opacity effects \citep{1996ApJ...457...604B,2011A&A...533..A111M}. These slight PA differences were considered and interpolated to the epochs without observation at 8.4\,GHz. We assembled all the PAs of VLBI observations from 1981 to 2022 at different frequencies to that of 8.4\,GHz, and show the results in Figure \ref{fig:fig1}A. Six VLBI epochs at 8.4\,GHZ are also shown in Fig. \ref{fig:fig1}D, where the jet wobbling can be seen around a projected on-sky axis (shown in red). There is a hint of periodic behavior, which can be fitted using a sinusoidal model.

\subsection{Light curves}
\begin{figure}[h!]
\centering
\includegraphics[width=3.4in]{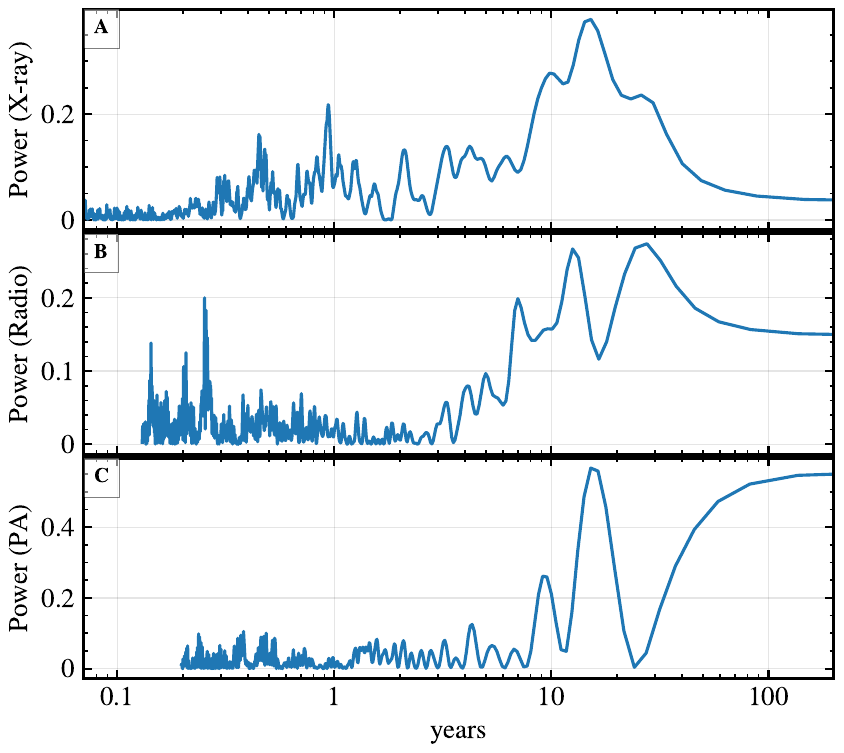}
\caption{Lomb-Scargle periodograms (LSP) of time series of X-ray (upper panel), radio fluxes (middle panel) and PAs (lower panel) for M81*. A peak at $\sim$15.2 years can be detected in the LSP of PAs and the X-ray light curve, respectively. There are two comparable peaks at 27.6 and 12.5 years in the radio LSP. All these time series measurements suggest a periodic mechanism in the central region.} \label{fig:fig5}
\end{figure}

The radio light curve is assembled to the flux densities at 8.4\,GHz or interpolated at 8.4\,GHz with a spectral index if the flux density measurements other than that at 8.4\,GHz are available. The averaged spectral index $\alpha$ of M81 core is $-0.11$, $F_{\upsilon}(\upsilon)\propto\upsilon^{-\alpha}$. It appears that there are repeated outbursts in the radio and X-ray light curves (\autoref{fig:fig1}). We search the periodicity in the time series of PA by using the Lomb-Scargle periodogram (LSP) calculated by \texttt{astropy} \citep{2013A&A...558A..33A,2018AJ....156..123A,2022ApJ...935..167A}. There is an obvious peak at 15.2 years (\autoref{fig:fig5}C), which indicates that the PA shows a periodic variation with $\sim$ 15 years. We also detect a peak at $\sim$ 15.2 years in the LSP of X-ray light curve (\autoref{fig:fig5}A), which is consistent with that of PA. Meanwhile, there are two comparable peaks at 27.6 and 12.5 years in the radio LSP (\autoref{fig:fig5}B). The differences of periods between radio and X-ray light curves should be due to the sparse sampling in the 1980s. The measuring of the periodicity would be improved with simultaneously intensive VLBI and high-energy monitoring in the future.

\section{Supermassive black hole binary model} 
\begin{figure}[h!]
\centering
\includegraphics[width=3.4in]{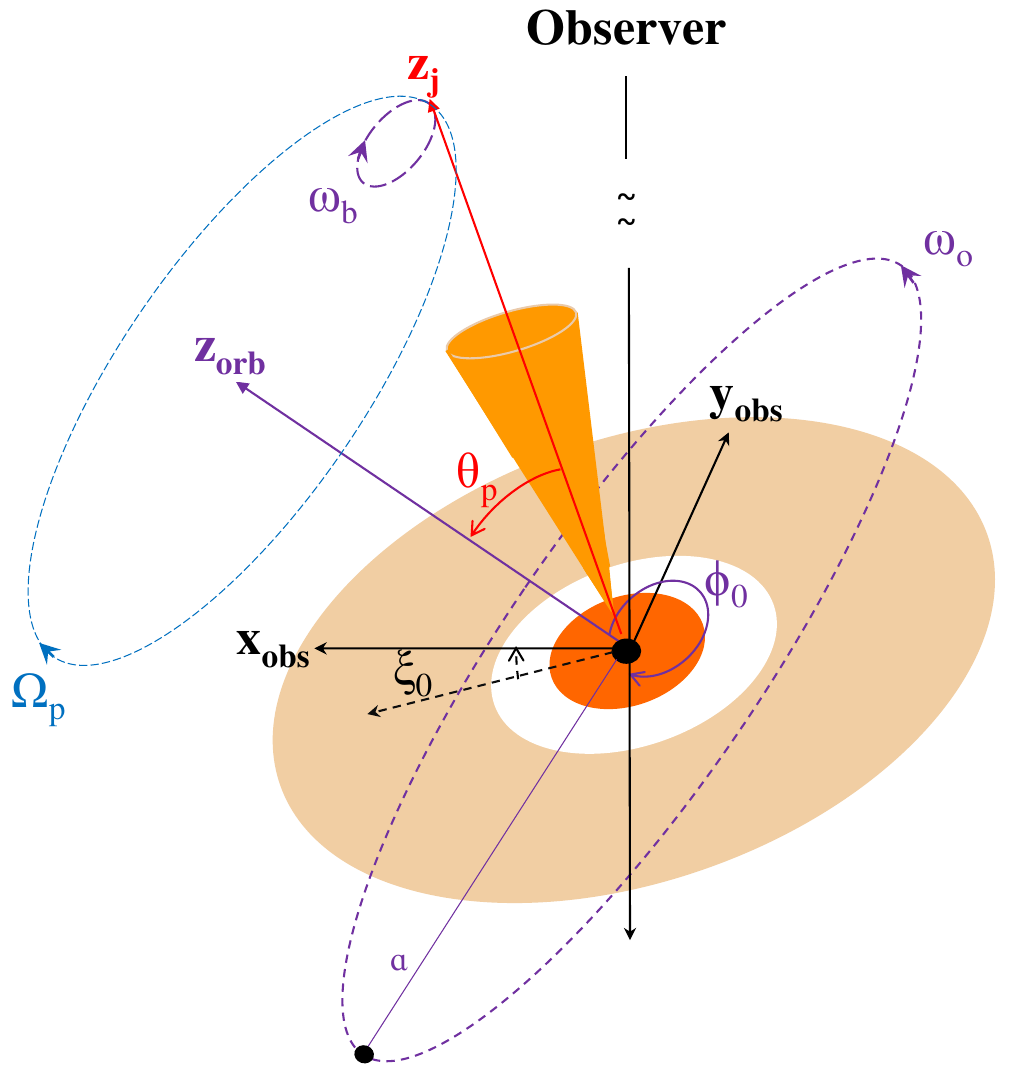}
\caption{Schematic representation of the black hole binary model in M81*. The secondary is orbiting around the primary black hole with an angular velocity of $\omega_o$ and a separation $a$. The orbital plane is misaligned $\theta_p$ to the disk plane of primary black hole. $\xi_0$ and $\phi_o$ are the projection of the orbital axis $\mathrm{\mathbf{z}}_\mathrm{{orb}}$ in the sky and its inclined angle to the LOS, respectively. The disk-jet of primary black hole undergoes short term ($2\pi/\omega_b$) nodding and wobbling (smaller purple ellipse), superimposing a long-term ($2\pi/\Omega_p$) precession (blue ellipse).} \label{fig:fig2}
\end{figure}

In a binary black hole system, where the disk plane is misaligned with respect to the orbital plane, the jet of the primary black hole is expected to undergo jet wobbling and nodding besides of the jet precession (see Fig. \ref{fig:fig2}), due to the oscillating torque from an orbital companion \citep{1982ApJ...260..780K,2000MNRAS.317..773B}. The wobble is along the precession direction and its rate $\omega_b$ is approximately twice of the orbital rate ($\omega_o$), while the amplitude in radians is roughly $\Omega_p/2\omega_o$, where $\Omega_p$ is the precession rate $\Omega_p\approx0.05\omega_o$. The nodding period is the same but the amplitude is multiplied by a factor of $\tan\theta_p$, where $\theta_p$ is the inclined angle of the orbit plane to the disk plane. 

A series of rotations are used to transform the motion of the jet vector $\mathbf{n_j}=(0, 0, 1)$ to the observer’s frame. All the rotations are in right handed coordinate systems. The first step is to be parallel to the orbital axis. $\mathbf{n_{orb}}$ in the orbital frame can be calculated using the following equations

\begin{equation}
\mathbf{n_{orb}} = \mathbf{R}_z(\theta_{wob})\mathbf{R}_y(\theta_{nod})\mathbf{n_j},
\end{equation}

\noindent where $\theta_{nod}=\theta_p+\tan(\theta_p)\frac{\Omega_p}{2(\omega_o-\Omega_p)}*\cos2((\omega_o-\Omega_p)(t-t_0)-\eta_p)$ and $\theta_{wob}=\Omega_p(t-t_0)+\frac{\Omega_p}{2(\omega_o-\Omega_p)}*\sin2((\omega_o-\Omega_p)(t-t_0)-\eta_p)+\eta_p$, $t_0$ is set to 1980 as the reference time here. Two more rotations are applied to transform to the observer's frame.

\begin{equation}
\mathbf{n_{obs}} = \mathbf{R}_z(\xi_0)\mathbf{R}_y(\phi_o)\mathbf{n_{orb}},
\end{equation}

\noindent where $\xi_0$ and $\phi_o$ is the projection of the orbital axis in the sky and its inclined angle to the LOS, respectively. 

\subsection{Bayesian analysis}\label{subsec:MCMC} 
\begin{figure}[h!]
\centering
\includegraphics[width=3.4in]{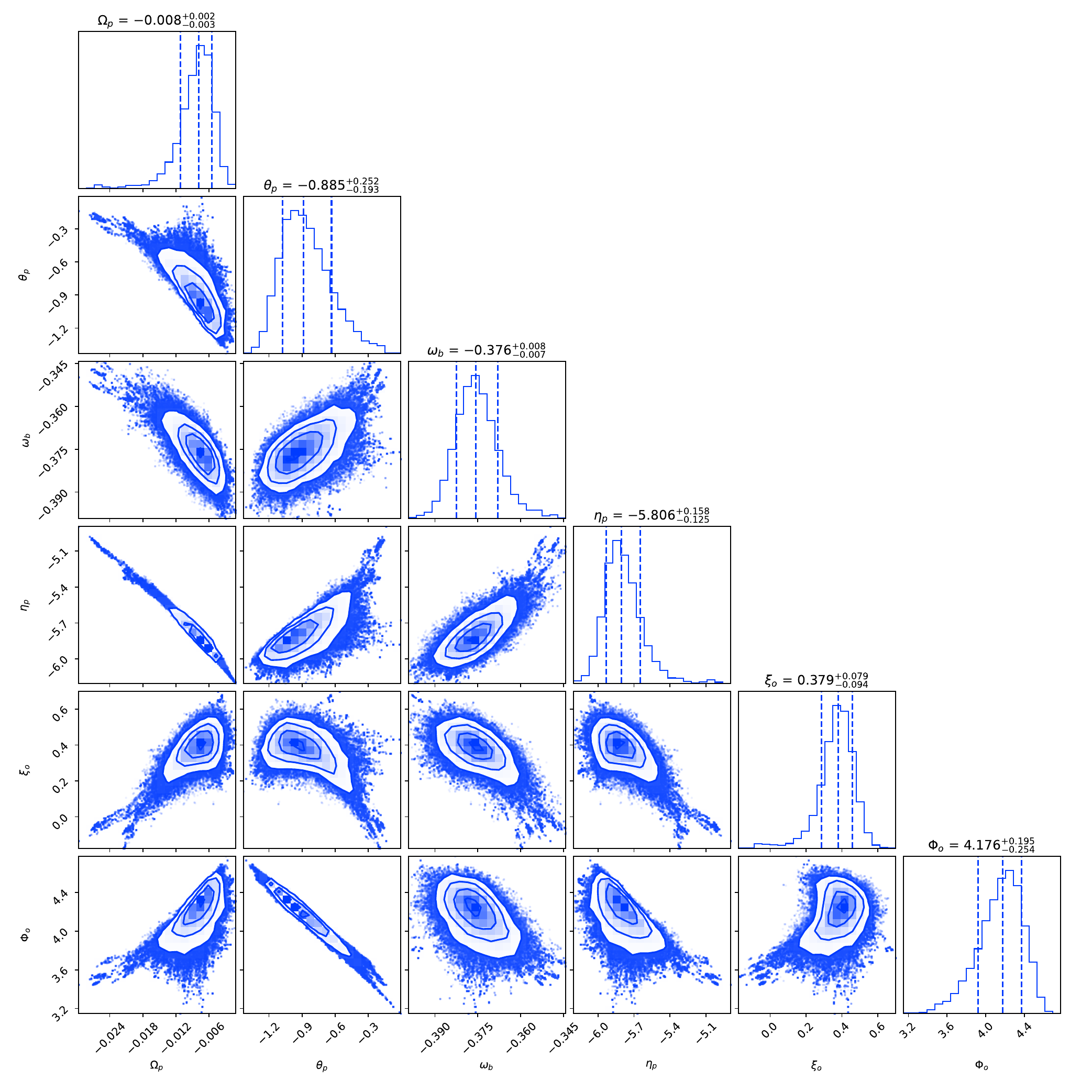}

\caption{Corner plot of the MCMC fitting to a binary black-hole model in M81*. The six parameters in the binary black hole model $\Omega_p$, $\theta_p$, $\omega_b$, $\eta_p$, $\xi_0$, and $\phi_o$ is the precession rate, the inclined angle of the orbit plane to the disk plane, the wobbling and nodding rate, the initial projection angle of jet on the orbital plane at $t_0$, the projection of the orbital axis in the sky and its inclined angle to the LOS, respectively.} \label{fig:fig3}
\end{figure}

The Bayesian analysis based on Markov Chain Monte Carlo (MCMC) is one of the most effective approaches to infer posterior distributions of model parameters, which describe the data while marginalizing over nuisance parameters and taking into account systematic uncertainties. It becomes a ubiquitous technique in astronomy \citep{2017ARA&A..55..213S}. A Bayesian analysis using an MCMC algorithm \citep{2013PASP..125..306F} is performed with six parameters ($\Omega_p$, $\theta_p$, $\omega_b$, $\eta_p$, $\xi_0$, and $\phi_o$) in the binary black hole model. The PA likelihood is assumed to be Gaussian, $-\ln\mathcal{L}_{PA}=\sum_{i}\frac{(PA(t_i)-MOD(t_i))^2}{2\sigma_i^2}$. In addition to the PA measurements, we consider joint constraints with the jet viewing angles  $\theta_{v,2011}<56^\circ$ in 2011 \citep{2016NatPh...12..772K} and  $\theta_{v,2018}<15^\circ$ in 2018 (Wang et al. in prep). The corner plot of MCMC fitting results is presented in Fig. \ref{fig:fig3}. All the results (given with $1\sigma$ uncertainties) are presented here. The jet wobble and precession model of a SMBHB fits the evolution of PA well with a reduced $\chi^2$ equal to 1.21 (the red line in \autoref{fig:fig1}A). 
The MCMC results tell the jet is wobbling every $\sim16.7$ years, and precessing with an additional long period $\sim785$ years. The binary model prefers an orbit highly inclined or misaligned to the inner accretion disk with $\theta_p={-54^{+10}_{-15}}^{\circ}$. 

\subsection{Repeated outbursts}\label{subsec:LC}
In the relativistic beaming theory, the observed and rest-frame flux densities are linked by $F_\upsilon(\upsilon)=\delta^{2+\alpha}F^{\prime}_{\upsilon^{\prime}}(\upsilon)$, where the Doppler factor $\delta=[\gamma(1-\beta \cos\theta_v)]^{-1}$ and the Lorentz factor $\gamma={(1-\beta^2)}^{-1/2}$ \citep{2017Natur.552..374R}. We use a bulk velocity $\beta=0.71$ in the jet core given the apparent speed of 0.51c \citep{2016NatPh...12..772K} and the viewing angle $12.8^\circ$ from fitting results. The averaged spectral index and the Doppler factor incorporating with the fitted viewing angle $\theta_v$ are applied for the radio light curve. The X-ray light curve (Fig \ref{fig:fig1}C) is also compensated the beaming effect due to the changes of viewing angles. The detection of relatively slow apparent speeds of knots ($\sim0.5\,c$ near the core in 2011 and $\sim0.1\,c$ down the jet in 2018) indicates the Doppler beaming effect in the jet should be small. The fitted descending viewing angles are not responsible for the repeated outbursts in the light curves. The LSP results of original light curves and those corrected the beaming effect are almost identical. We suppose a compact object circling the primary black hole in a Keplerian orbit. When the compact object goes through the accretion disk, the mass accretion rates are raised and therefore cause outburst periodically. A binary system can enhance the mass accretion rate twice per binary orbit \citep{2013PASJ...65...86H,2014PhRvD..89f4060G}. 

\section{discussions} \label{sec:dis}
\subsection{Implications of the Observational Results}\label{subsec:Imp}
Implied from the SMBHB model as shown in Fig.\ref{fig:fig2}, the troughs of fitting viewing angles in each wobbling period occur when the secondary black hole goes through the accretion disk. The scenario is similar to the most promising SMBHB candidate OJ\,287, although the origin of flares is proposed to be impact flashes \citep{2008Natur.452..851V} or precession-induced \citep{2023ApJ...951..106B}. The occurence time of troughs in viewing angles are consistent with those of flux peaks in light curves. Interestingly, the intervals among three peaks in the light curves are similar but not identical. That means the two nodes in an orbit are not symmetric and the orbit period should not be exactly twice of the jet wobbling period. We adopt the time interval of $29.7\pm0.4$\,yr between the peaks in 1985 and 2015 as the orbit period, which corresponds to an orbit separation $a\sim0.02$\,pc ($\sim1$\,mas at the distance $\sim4$\,Mpc) via the Kepler’s third law. The post-Newtonian parameter \citep{2011PNAS...108..5938W}, $\varepsilon=GMc^{-2}r^{-1}$ is about $1.5\times10^{-4}$. A value of $10^{-8}<\varepsilon<10^{-3}$ confirms that it is a bounded wide binary system where dynamical friction is dominant at this stage. A mass ratio of the companion to the primary black hole, $0.07<q<0.10$, could be constrained through its relations to the outer disk radius and the precession period, the equations (14) and (16) in \citet{2018MNRAS...478..3199B}, respectively. A small mass ratio indicates the companion is probably radio quiet due to no normal accretion disk at most of its orbital phase. The gravitational wave (GW) strain $h$, as expressed by the equation (6) in \cite{2020ApJ...889...79Y}, is about $3\times10^{-17}$, which is more than one order lower below the detection limit of the strain-sensitive GW detector, international pulsar timing arrays \citep{2022MNRAS.510.4873A,2023ApJ...951L...8A,2023A&A...678..A50E,2023ApJ...951L...6R,2023RAA....23g5024X}. The parameters of the SMBHB model are listed in Table \ref{tab:tab2}. We estimate the separation when gravitational radiation drives the binary to merger, it is about 0.004\,pc. The coalescence timescale in the GW-dominated regime is $\sim$3.9 million years \citep{2010PhRvD..81d7503L}. While the inspiral from 0.02\,pc to 0.004\,pc could be driven by the gas-assisted orbital contraction, which depends on the accretion dynamics of circumbinary disk and the binary system itself. This is related to the interesting ``final-parsec problem" \citep{1980Natur...287..307B, 2023ARA&A..61..517L} and should be investigated with further studies. 

Regarding the jet launching of the primary black hole, two widely adopted scenarios have been proposed. The outer disk radius of inner accretion disk is $\sim1000\,r_g$, far less than the orbital separation $\sim6000\,r_g$, which suggests that the secondary black hole is not significantly affecting the Blandford-Payne (BP) jet launched by the primary disk \citep{1982MNRAS.199..883B}. Blandford-Znajek (BZ) jet is coupled to the spin of black hole \citep{1977MNRAS.179..433B}, it is thus more independent to the companion black hole. Current data can not distinguish the scenario of BP or BZ. In the case of mass ratios $\ll$1, the accretion and variability are dominated by the accretion flow onto the primary \citep{2014PhRvD..89f4060G}. As proposed in Section 4, the jet precession of primary black hole is due to the misaligned disk to orbital plane \citep{1982ApJ...260..780K,2000MNRAS.317..773B, 2023ARA&A..61..517L}. We can not distinguish the scenario of disk-jet precession or precession of disk-jet and its central black hole as a whole either.

\subsection{Other Possible Scenarios}
A jet launched by a tilted accretion disk of spinning SMBH undergoes Lense-Thirring (LT) precession \citep{2007ApJ...668..417F,2018MNRAS...474..L81L}. The sinusoidal changes of PA could be related to a precessing jet. The long-term precession modulation ($\Omega_p$) could be contributed by an outer elliptical-disk \citep{1996AJ....111.1901B,2003ApJ...598..956S}. Including $\Omega_p$, there are six parameters in the LT precession model ($\omega_{LT}$, $\theta_{LT}$, $\Omega_p$, $\eta_0$, $t_0$, and $\Phi_o$). We also perform a Bayesian analysis using an MCMC algorithm. We consider joint constraints of the PA measurements together with the jet viewing angles  $\theta_{v,2011}<56^\circ$ in 2011 and  $\theta_{v,2018}<15^\circ$ in 2018, as well as a further constraint is given by the inclination angle $14\pm2^{\circ}$ of the LOS to the normal direction of the gaseous disk \citep{2003AJ...125...1226D}, $\Phi_o=194\pm2^{\circ}$ as the LOS is in the opposite direction to the observer. The fitting results indicate a $17.5\pm1.5$ year periodic LT precession $({2\pi}/{\omega_{LT}})$, superimposed to a long-term rotation period $({2\pi}/{\Omega_{p}})$ of $628.3\pm63.0$ years. The tilted angle of the inner thick disk is $\theta_{LT}=1.7\pm0.6$ degrees.

A small tilt angle implies the disk precession by LT should have entered a stable stage. According to the equation (43) in \citet{2007ApJ...668..417F}, the short period of LT precession indicates a truncated inner disk at $160-170  r_g$, incorporating a black hole mass of $7\times10^7M_\odot$ and a spinning parameter $a_{\bullet}=0.9$.  However, the radius of truncated inner disk by LT is larger than a typical value of several tens $r_g$ near the black hole horizon. And forming an outer elliptical-disk implies a possible companion. Additionally, the repeated outbursts could not be explained by the Doppler boosting of a reduced viewing angle in the case of LT precession. Using the fitted viewing angles, the peaks in the light curves (Fig. \ref{fig:fig1}B) should be produced with a moderate Lorentz factor $\Gamma\sim12.9$, a relativistic velocity $\beta=0.997$ of the emitting source in units of the speed of light. This is inconsistent with the slow motions of knots detected in the jet of M81*.

Lense-Thirring precession of the SMBH itself has a normal period of millions of years. Otherwise, the spinning parameter of M81* black hole would have to be very small ($\sim10^{-12}$ for a timescale of the order of a decade), according to equation (10) in \cite{2018MNRAS...478..3199B} with
a viscosity parameter $\alpha_{vis}$ = 0.1. The scenario of the SMBH precession is thus rather implausible. 

\section{Conclusions} \label{sec:sum}
The jet precession in M81* is investigated with four-decade data thoroughly both in radio and X-ray observations. Owning to the proximity of M81*, the high resolution VLBI approach the inner core region adjacent to the black hole, and reveals that the M81* jet is undergoing a short period of jet wobbling $\sim16.7$ years, superposing a long-term precession at a time-scale of several hundred years. Combining the PA evolution and the light curves, this can be best explained by oscillations of a secondary black hole in a binary system, rather than black-hole precession or tilted disk-jet precession due to LT effect. The binary system has an orbital separation of $\sim 0.02$\,pc and an orbit period of $\sim30$\,yr. This makes M81* as one of the closest SMBHB candidate and a potential target to resolve the ``final parsec problem". Nevertheless, we notice that similar claims about existing binary black holes have been made before but turn out to be challenging, as evidenced by the detection of the offset of VLBI component not associated with the VLBI core \citep{2013A&A...557A..85R} or possibly dual inverted-spectrum radio cores \citep{2017NatAs...1..727K}, as well as the extensive literature on OJ287 and its multiple unsuccessful predictions. Some recent evidences even contradict the existence of supermassive binary black holes, e.g. the SMBHB candidate J1502SE/SW are double hotspots \citep{2014ApJ...792L...8W}, PSO\,J334 is most likely a jetted active galactic nucleus with a single SMBH \citep{2023A&A...677A...1B}, against the binary scenario evidenced by periodic variations in its optical light curve. Future intensive contemporary VLBI and high-energy monitoring campaigns are necessary to further verify the SMBHB scenario in M81*. \\

\begin{acknowledgments}
The authors thank the anonymous referee for very constructive suggestions in improving the manuscript. WJ thanks Fangzheng Shi, Zhiyuan Li for useful discussions.
This work was supported in part by the National Natural Science Foundation of China (Grant No. 12173074, 11803071). IMV thanks the Generalitat Valenciana for funding in the frame of the CIDEGENT/2018/021 and ASFAE/2022/018 Projects, the MICINN Research Project PID2019-108995GB-C22, and the Astrophysics and High Energy Physics programme by MCIN, with funding from European Union NextGenerationEU (PRTR-C17I1). ZY is supported by the Natural Science Foundation of China (grants U1838203,U1938114), the Youth Innovation Promotion Association of CAS (id 2020265) and funds for key programs of Shanghai astronomical observatory. YL is supported in part by the Natural Science Foundation of Shanghai (grant NO. 23ZR1473700).
We are grateful to all staff members at Effelsberg-100m, EVN, GBT, KaVA, Sheshan \& Tianma-65m, VLA, and VLBA, who helped to operate the array and to correlate the data. VLBA, VLA, GBT, and the data archive system are operated by the National Radio Astronomy Observatory, which is a facility of the National Science Foundation operated under cooperative agreement by Associated Universities, Inc. The European VLBI Network is a joint facility of independent European, African, Asian, and North American radio astronomy institutes. The KVN is a facility operated by the Korea Astronomy and Space Science Institute. VERA is a facility operated by National Astronomical Observatory of Japan in collaboration with Japanese universities. 

\textit{Software:} \tt AIPS, \tt Astropy, \tt CASA, \tt DIFMAP, \tt PyXspec, \tt UltraNest
\end{acknowledgments}



\clearpage

\begin{longtable}{p{2.5cm}p{4.0cm}p{3.0cm}p{2.5cm}p{2.5cm}clcll}
\caption{Summary of VLBI/VLA observations on M81* }
\label{tab:tab1} \\
\hline {Epoch} & {Array \& telescopes}& {Frequency}& {$F_{8.4\,\mathrm{GHz}}$/error}& {PA/error} \\ 
{yyyy-mm-dd} & {}  & {GHz}& {mJy/mJy}& {deg./deg.} \\ 
\hline
\endfirsthead
\multicolumn{6}{l}%
{{\bfseries \tablename\ \thetable{} -- continued from previous page}} \\
\hline 
{Epoch}  & {Array\& telescopes} & {Frequency} & {$F_{8.4\,\mathrm{GHz}}$/error} & PA \\ 
{yyyy-mm-dd} & {} & {GHz}& {mJy/mJy}& {deg./deg.} \\ \hline
\endhead

\endfoot

\hline \hline
\endlastfoot
\multicolumn{6}{c}{VLBI } \\
\hline
2005-07-13  &{VLBA, EF}  & 8.4    & 64.1/1.0 & 62.7/3.6  \\
2006-06-16  &VLBA, EVN, GB, VLA  & 5.0  & 74.6/0.8 & 55.1/7.2 \\
2007-01-26  &VLBA, EF  & 8.4/15 & 70.5/0.8 & 53.0/3.0 \\
2007-11-03  &VLBA, EVN, GB, VLA  & 5.0  & 74.1/0.8 & 53.0/7.2 \\
2008-05-02  &VLBA, EF &  8.4/15/22  & 91.5/1.2 & 62.5/2.4 \\       
2009-04-10  &VLBA, EF &  8.4/15/22  & 102.0/0.8 & 63.2/3.0  \\   
2009-04-30  &VLBA &  5.0/8.4/15   & 107.2/1.2 & 58.0/6.0    \\   
2009-10-04  &VLBA, EF &  5.0/8.4/15 & 93.9/0.8 & 61.6/3.0    \\   
2009-12-04  &VLBA, EF &  5.0/8.4  & 72.7/1.2 & 61.7/3.0    \\   
2010-02-06  &VLBA, EF &  5.0/8.4  & 86.7/0.9 & 60.4/1.8    \\   
2010-03-20  &VLBA, EVN &  5.0 & 83.6/0.9 & 56.4/7.2      \\   
2010-05-04  &VLBA, EF &  5.0/8.4  & 60.6/0.9 & -    \\   
2010-05-29  &VLBA, EVN &  5.0  & 64.6/0.9 & 59.6/7.2    \\  
2010-07-02  &VLBA, EF &  5.0/8.4 & 90.7/1.2 & -     \\  
2010-09-05  &VLBA, EF &  5.0/8.4 & 92.5/1.2 & -   \\  
2010-11-19  &VLBA, EF &  5.0/8.4 & 111.7/0.9 & 52.6/1.8   \\   
2011-03-18  &VLBA, EF & 5.0/8.4 & 129.9/0.9 & 55.8/1.8    \\ 
2011-05-12  &VLBA, EF & 5.0/8.4 & 95.0/0.9 & 58.6/1.8    \\ 
2011-08-14  &VLBA &  8.4  & 107.3/1.0 & 59.1/3.0     \\ 
2011-09-17  &VLBA & 8.4  & 143.4/1.0 & 59.4/3.0    \\ 
2011-12-19  &VLBA, EF &  5.0/8.4 & 95.0/0.9 & 61.2/1.8    \\ 
2012-01-16  &VLBA & 15/22  & 77.2/1.2 & 61.4/3.0    \\ 
2012-08-25  &VLBA, EF, GB &  5.0/8.4 & 77.9/1.3 & 66.3/3.0  \\ 
2013-01-28  &VLBA, EF, GB &  5.0/8.4 &72.3/1.2 & 62.3/2.4  \\
2015-03-20   &KaVA &  22   & 181.8/1.0 & 64.7/9.0 \\
2015-03-21   &KaVA &  43 & 181.8/2.0 & 64.7/9.0    \\
2015-11-21  &VLBA, EF, GB & 22 & 92.8/1.2 & 61.5/2.4   \\ 
2016-02-27   &VLBA &  43   & 83.0/1.4 & 69.1/4.5  \\ 
2016-12-15  &VLBA, EF, GB &  22 & 77.0/1.2 & 72.5/2.4    \\ 
2016-12-17  &VLBA, EF, GB &  22 & 93.3/1.3 & 70.3/2.7  \\ 
2017-12-17  &VLBA, EF, GB &  22 &75.9/1.3 & 80.8/2.4  \\ 
2017-12-18  &VLBA, EF, GB & 22 &73.1/1.0 &  78.9/2.7  \\ 
2018-04-22  &VLBA & 43 & 125.6/1.0 & 83.5/4.5  \\
2018-05-14  &VLBA & 8.4/22/43 & 127.1/8.0 & 83.3/3.9 \\
2018-06-10  &VLBA & 8.4/22/43 & 91.6/0.9 & 84.7/3.0 \\
2019-11-02  &VLBA & 22 & 132.2/3.0 & 78.3/4.5     \\
2019-11-04  &VLBA & 5.0/8.4  & 146.0/1.0 & 76.5/3.0 \\
2021-01-12  &VLBA & 22 & 78.7/1.1 & 80.1/3.0    \\
2022-02-23  &VLBA, SH, T6 &  8.4/22 & 124.1/10.0 & 74.4/3.0 \\
\hline
\multicolumn{6}{c}{VLA} \\
\hline
1980-11-07  &VLA&5.0 & 64.2 &-\\
1981-02-07  &VLA& 5.0 & 95.7 &-\\
1981-08-16  &VLA& 5.0 & 94.8 &-\\
1982-05-31  &VLA& 5.0 & 85.1 &-\\
1982-07-02  &VLA& 5.0/15 & 88.5 &-\\
1982-10-25  &VLA& 5.0/15 & 86.2 &-\\
1982-11-06  &VLA& 5.0 & 79.6 &-\\
1982-12-08  &VLA& 5.0/15 & 75.5 &-\\
1983-01-20  &VLA& 5.0/15 & 90.1 &-\\
1983-01-28  &VLA& 5.0 & 79.6 &-\\
1983-07-31  &VLA& 5.0/15 & 65.8 &-\\
1983-11-30  &VLA& 5.0/15 & 70.1 &-\\
1983-12-01  &VLA& 5.0 & 73.3 &-\\
1984-03-01  &VLA& 5.0/15 & 81.5 &-\\
1984-05-31  &VLA& 5.0/15 & 92.2 &-\\
1984-06-26  &VLA& 5.0/15 & 75.1 &-\\
1984-11-10  &VLA& 5.0/15 & 79.2 &-\\
1985-02-09  &VLA& 5.0/15 & 97.4 &-\\
1985-02-13  &VLA& 5.0 & 110.9 &-\\
1985-03-25  &VLA& 5.0/15 & 111.8 &-\\
1985-04-13  &VLA& 5.0/15 & 124.1 &-\\
1985-05-04  &VLA& 5.0/15 & 136.1 &-\\
1985-06-21  &VLA& 5.0/15 & 107.3 &-\\
1988-01-17  &VLA& 5.0 & 108.4 &-\\
1988-06-09  &VLA& 5.0 & 141.1 &-\\
1988-16-09  &VLA& 5.0/15 & 125.0 &-\\
1989-03-13  &VLA& 5.0 & 128.5 &-\\
1990-07-29  &VLA& 5.0 & 167.5 &-\\
1991-12-06  &VLA& 5.0 & 109.6 &-\\
1998-11-13  &VLA& 5.0/8.4 & 233.7 &-\\
1998-11-20  &VLA& 5.0/8.4 & 239.6 &-\\
1998-12-07  &VLA& 5.0 & 245.0 &-\\
1999-06-06  &VLA& 8.4/22 & 203.1 &-\\
1999-09-05  &VLA& 5.0/8.4 & 287.6 &-\\
1999-11-23  &VLA& 5.0 & 253.4 &-\\
2000-09-27  &VLA& 5.0/8.4 & 187.8 &-\\
2014-01-06	&VLA&	5.0 &  82.9 &-\\
2014-03-04	&VLA&8.4/22.9 & 118.5 &- \\
2014-10-23	&VLA&5.8/8.4 & 160.1 &-  \\
2015-01-03	&VLA&5.8 & 249.4 &- \\
2015-03-27	&VLA&5.8 & 171.0 &- \\
\end{longtable}
\tablecomments{Col.(1) Observing epoch;  Col.(2) Array and telescopes used. VLBA has an interferometric beam size of $1.4\,\mathrm{mas}\times1.4\,\mathrm{mas}$, $0.8\,\mathrm{mas}\times0.8\,\mathrm{mas}$, $0.45\,\mathrm{mas}\times0.45\,\mathrm{mas}$, $0.3\,\mathrm{mas}\times0.3\,\mathrm{mas}$, and $0.15\,\mathrm{mas}\times0.15\,\mathrm{mas}$ at 5.0\,GHz, 8.4\,GHz, 15\,GHz, 22\,GHz, and 43\,GHz, respectively. The resolution can be 1.4-1.5 times higher if adding European or Chinese telescopes on M81*; Col.(3) Observing frequency bands; Col.(4) Flux densities $F_{8.4\,\mathrm{GHz}}$ at 8.4\,GHz or interpolated to 8.4\,GHz with a spectral index. The error of VLA flux density measurement is 10\% of its magnitude; Col.(5) PAs at 8.4\,GHz or interpolated to 8.4\,GHz with averaged frequency differences.}
\clearpage

\linespread{0.5}
\begin{deluxetable}{cll}
\tablecaption{Parameters of SMBHB system in M81* \label{tab:tab2}}
\tablenum{2}
\tablehead{
\colhead{Notation} & \colhead{Definition} & \colhead{Quantity} 
}
\startdata
$P_o$	& Orbit period	& $29.7\pm0.4$ yr \\
$P_p$	& Precession period	& $571-785$\,yr \\
$\theta_p$	&Inclination of the orbital plane to the disk plane& ${-54^{+10}_{-15}}^\circ$ \\
$a$	& Orbital separation& $\sim0.02$\,pc ($\sim1.0$\,mas) \\
M	& Total mass	& $\sim7\times10^7 M_\odot$ \\
q	& Mass ratio of companion to the primary black hole	& $0.07-0.10$ \\
$f_g$	& Gravitational wave frequency	& $\sim2.0\times10^{-9}$\,Hz \\
$h$	& Gravitational wave strain & $\sim3.0\times10^{-17}$ \\
\enddata
\end{deluxetable}
\clearpage


\end{document}